\documentclass[twocolumn,aps,superscriptaddress,nofootinbib]{revtex4-1}

\usepackage{graphicx}
\usepackage{braket} 
\usepackage{float}%
 \usepackage{amsmath}
\usepackage{amssymb}
\usepackage{amsfonts}
 \usepackage{bbold} 
 
 \usepackage[normalem]{ulem} 
 
\usepackage{slashed} 

\usepackage[colorlinks]{hyperref}  

\newcommand{\sP}{\slashed{p}}
\newcommand{\spk}{\slashed{p}_k}
\newcommand{\sq}{\slashed{q}}
\newcommand{\sPr}{\slashed{P}}

\newcommand{\sPxi}{\slashed{p}_\Xi}

\newcommand{\ra}{\rangle}
\newcommand{\la}{\langle}

\graphicspath{{./plots/}}

\begin{document}

\title{Weak production of strange $\Xi$ baryons off the nucleon}

\author{M. Rafi \surname{Alam}}
\email{rafi.alam.amu@gmail.com}
\affiliation{Department of Physics, Aligarh Muslim University, Aligarh-202 002, India}
\affiliation{Departamento de F\'\i sica Te\'orica and IFIC, Centro Mixto
Universidad de Valencia-CSIC, Institutos de Investigaci\'on de
Paterna, E-46071 Valencia, Spain}
\author{I. Ruiz  \surname{Simo}}
\email{ruizsig@ugr.es}
\affiliation{Departamento de F\'\i sica At\'omica, Molecular y Nuclear and Instituto Carlos I de F\'\i sica Te\'orica y Computacional, Universidad de Granada,
E-18071 Granada, Spain}

\begin{abstract} 
The charged current
Cabibbo-supressed associated 
$K \Xi$ production off the nucleon 
induced by anti-neutrinos 
is studied 
at low and intermediate energies. 
The non-resonant terms are obtained 
using a microscopical model based 
on the SU(3) chiral Lagrangian. 
The basic parameters of the model 
are $f_\pi$, the pion decay 
constant, Cabibbo's angle, the 
proton and neutron magnetic moments 
and the axial vector coupling 
constants for the baryons octet, D 
and F, that are obtained from the 
analysis of the semileptonic decays 
of neutron and hyperons. 
In addition, we also consider 
$\Sigma^\ast(1385)$ resonance, 
which can decay in $K \Xi$ final 
state when this channel is open. 
The studied mechanism is the 
prime source of $\Xi$ 
production at 
anti-neutrino energies around $2$
GeV and the calculated cross 
sections at 
these energies can be measured at 
the current and future neutrino 
experiments.
\end{abstract}

\maketitle

\section{Introduction}

With the recent advancements in 
the experimental facilities, 
it is now 
possible to test several reactions 
which were lately considered not 
to be observable. One of such 
interactions is 
the strange particle production. 
In last decade there were several 
attempts to explore the 
strangeness physics in weak 
sector~\cite{Alvarez-Ruso:2014bla,
Katori:2016yel}. The quasi-elastic 
production of 
hyperons~\cite{Singh:2006xp,
Alam:2014bya,Wu:2013kla,
Sobczyk:2019uej} and their 
subsequent decay into 
pions~\cite{Alam:2013cra,
Fatima:2018wsy} have 
been recently  
analyzed around $\sim 1$ GeV  
anti-neutrino energies. 
Experimental efforts are also made 
to observe such 
processes~\cite{Farooq:2016ito}. 
Moreover, recent experiments like 
MINER$\nu$A~\cite{Minerva}, MicroBooNE~\cite{Microboone},  NO$\nu$A~\cite{NOvA}
and ArgoNeut~\cite{Argoneut} are capable of 
detecting such processes with high 
statistics.
These channels are (or 
proposed to) being updated in the 
modern event generators like 
GENIE~\cite{Andreopoulos:2009rq}, 
NEUT~\cite{Hayato:2009zz}, 
NuWro~\cite{Nuwro} and 
GiBUU~\cite{GiBUU}. 
However, particles with higher 
strangeness content are still 
unexplored both theoretically as 
well as experimentally.  

In this work we examine
for the first time 
the (higher) strange 
particle production processes 
induced 
by charged current weak 
interactions. We focus on channels 
where $\Xi$ baryons are in the 
final state. Since strange quantum 
number ($S$) of $\Xi$  is $-2$, 
the quasi-elastic channels 
are ruled 
out\footnote{$|\Delta S| =2$ 
processes are highly suppressed.}. 
The next primary source of $\Xi$ 
production is the 
inelastic channel where $\Xi$ is  
accompanied by a $K$ meson.  The 
process may be identified as 
$|\Delta S|=1$ and therefore 
suppressed by $\sin^2 \theta_c$, 
with $\theta_c(\sim 13^\circ)$ 
being the Cabibbo angle. 
The other competing 
process would be $\Xi KK$
production, which would be 
kinematically suppressed at
the energies probed in this work
due to its higher production
threshold. 
Here we would like to 
emphasize that the $|\Delta S|=1$ 
associated $K \Xi$ production 
can only be initiated 
by anti-neutrinos and not by 
neutrinos. The neutrino 
channels are forbidden due to 
phenomenological rule that 
examines the allowed quark 
transitions and 
is known as the 
$\Delta S = \Delta Q$ rule.
Further, for an 
experiment which would be 
capable of doing 
semi-exclusive measurements 
for $K \Xi$ production, one 
should notice that the 
exclusion of neutrino 
induced processes leaves 
only $\Delta S=0$ 
associated $Y K$ (where
$Y$ is any hyperon with
$S=-1$)
processes for kaons in 
the final state.  
While single $(\bar K)K$ are produced via 
$|\Delta S|=1$ 
(anti)neutrino induced 
processes 
\cite{RafiAlam:2010kf,
Alam:2011xq}
and hence will not 
contaminate the 
anti-neutrino 
induced $K$ production. This 
implies that the 
mechanism described here
is well suited to study 
the (semi-)exclusive $\Xi$ 
production induced by 
anti-neutrinos.

In this work, we extend our model 
of $|\Delta S|=1$ single 
kaon/antikaon 
production~\cite{RafiAlam:2010kf,
Alam:2011xq} to $K \Xi$ associated 
$|\Delta S|=1$ production. The non 
resonant background (NRB) terms 
are obtained from the expansion of 
chiral Lagrangian. We have
improved our 
previous model by including the 
phenomenological transition form 
factors based on flavor SU(3) 
symmetry. Possible effects due to 
symmetry breaking are also 
explored.
Finally, due to the 
high invariant mass 
of the $K \Xi$ system, we have also
considered the possible effects of 
resonances.
However, since the production 
mechanism is driven by a 
$|\Delta S|=1$ weak charged current, 
only strange resonances (with
$S=-1$) can contribute.
Very little or no information is 
available about these resonances
and their transition form factors,
both from experimental as well as 
theoretical sides. In view of this 
we have considered the lowest lying
strange resonance,
$\Sigma^\ast(1385)$, and estimate 
the effects of this in the present 
model. 

The paper is organized as follows: 
In section~\ref{sec:model} we 
discuss briefly our model for 
non-resonant (NRB) and resonant 
mechanisms. 
We present our numerical 
results in 
section~\ref{sec:results},
where we also discuss the 
possible effects due to the 
SU(3) symmetry breaking and 
highlight the possible outcome of 
it by using an explicit
model for SU(3) breaking
\cite{Faessler:2008ix}, which
is sketched and summarized in
appendix \ref{app:SU3}. 
Finally, we conclude 
our findings in 
section~\ref{sec:conclusion}.

\section{Model}\label{sec:model}
At low neutrino energies, the 
first channel that could produce 
$\Xi$ particles in final state 
proceeds through
charged current $|\Delta S|=1$ 
mechanism and the reactions are:  
\begin{align}\label{eq:all_weak_ch}
    \bar \nu_\mu + p & \rightarrow \mu^+ + K^+ + \Xi^- \nonumber \\
    \bar \nu_\mu + p & \rightarrow \mu^+ + K^0 + \Xi^0  \\
    \bar \nu_\mu + n & \rightarrow \mu^+ + K^0 + \Xi^- . \nonumber    
\end{align}
The double differential cross 
section
in the Lab frame  
for~(\ref{eq:all_weak_ch}) may be 
written as
\begin{equation}\label{eq:diff_3b_lab_phsp}
\begin{aligned}
    \frac{d \sigma}{d W d Q^2 } &= 
    \frac{1}{32(2 \pi)^5}  \int d E_K 
    \, \frac{\pi W  }{E_\nu^2 M^2 |\vec q|}\\
    & \times \Theta(1-\cos^2\theta_0)\, 
    \int d \phi_K  \,
    \overline \sum \sum |{\cal M}|^2 , 
    \end{aligned}
\end{equation}
where $W=\sqrt{(p+q)^2}$ is the invariant
mass of the final $K\Xi$ hadronic state, 
$Q^2=|\vec{q}|^2-(q^{0})^2$ is the positive
four-momentum transfer, and $|\vec{q}|$ and $q^0$
are the three-momentum and energy transfer
to the initial nucleon of mass $M$, respectively.

Finally,
\begin{equation}\label{cosine_sol}
    \cos\theta_0=\frac{M^2_\Xi+
    |\vec q|^2+|\vec{p}_K|^2-
    (M+q^0-E_K)^2}{2\,|\vec q|\, |\vec{p}_K|}
\end{equation}
is the cosine of the polar angle between
the kaon three-momentum $\vec{p}_K$ 
and the momentum transfer
$\vec{q}$ that ensures energy conservation 
through the step function $\Theta$, and
the integration over final kaon kinematics is
performed over the energy of the final kaon
$E_K$ and over
the azimuthal angle $\phi_K$ 
between the reaction plane
(that formed by the momenta of the kaon and the
cascade hyperon) and the lepton scattering plane.

It is worth noticing that, in Eq. 
(\ref{eq:diff_3b_lab_phsp}), once the antineutrino
energy $E_\nu$ is given, fixing $W$ and $Q^2$
means to fix the energy transfer $q^0$ and the
three-momentum transfer $|\vec{q}|$ as well. 
Therefore, the cosine of the angle between
$\vec{p}_K$ and $\vec{q}$, Eq. 
(\ref{cosine_sol}), is only fixed in
the integral by the value of the kaon energy,
but it is otherwise independent on $\phi_K$.

The average and sum over
initial and final particles' spins of the
square of the transition matrix 
element ($\mathcal M$) is given by  
\begin{align}\label{eq:amp2_weak}
 \overline{\sum} \sum |\mathcal{M} |^2 =  \frac{G_F^2}{2}  L_{\mu \nu} H^{\mu \nu},
\end{align}
where $G_F$ is the Fermi coupling
constant. 
The leptonic tensor $L_{\mu \nu}$ is, 
\begin{equation}
    L_{\mu \nu}= 8 \left[  k_\mu  k^\prime_\nu  +  k_\nu  k^\prime_\mu  - g_{\mu \nu} k \cdot k^\prime 
    -  i \, \varepsilon_{\mu \nu \alpha \beta} k^\alpha {k^\prime \,}^\beta \right] \, ,
\end{equation}
where our convention for the Levi-Civita
tensor in four dimensions is 
$\varepsilon^{0123}=-\varepsilon_{0123}=1$, 
and $H^{\mu \nu}$ is the hadronic tensor 
expressed in terms of hadronic current $J^{\mu}$, 
\begin{equation} 
\begin{aligned}
    H^{\mu \nu} &= \frac12 {\rm tr} \left[ J^\mu (\sP +M )  \tilde J^{\nu} 
    (\sPxi^\prime +M_\Xi ) \right]  \\
    \tilde J^{\nu} &=  \gamma^0 J^{\nu ^\dagger} \gamma^0
    \label{hadron-tensor}
    \end{aligned}
\end{equation}
The hadronic current in Eq.
(\ref{hadron-tensor}) corresponds
to the amputated amplitude 
(without Dirac spinors) obtained 
from Eqs. (\ref{eq:born_amp}) and 
(\ref{eq:res_amp}) of sections
\ref{NRB-section} and
\ref{Resonant-section} discussed
below, i.e, $j^\mu_{cc}|_{i}=
\bar{u}_{\Xi}(p^\prime_\Xi)J^\mu_i
u_{N}(p)$, and the total
hadron current is $J^\mu=\sum_{i} 
J^\mu_i$, where the index $i$ runs
over all the possible contributions (Feynman
diagrams)
that yield the same final hadron
state specified in Eqs. 
(\ref{eq:all_weak_ch}).

\begin{figure}
\begin{center}
\includegraphics[width=0.4\textwidth,height=.25\textwidth]{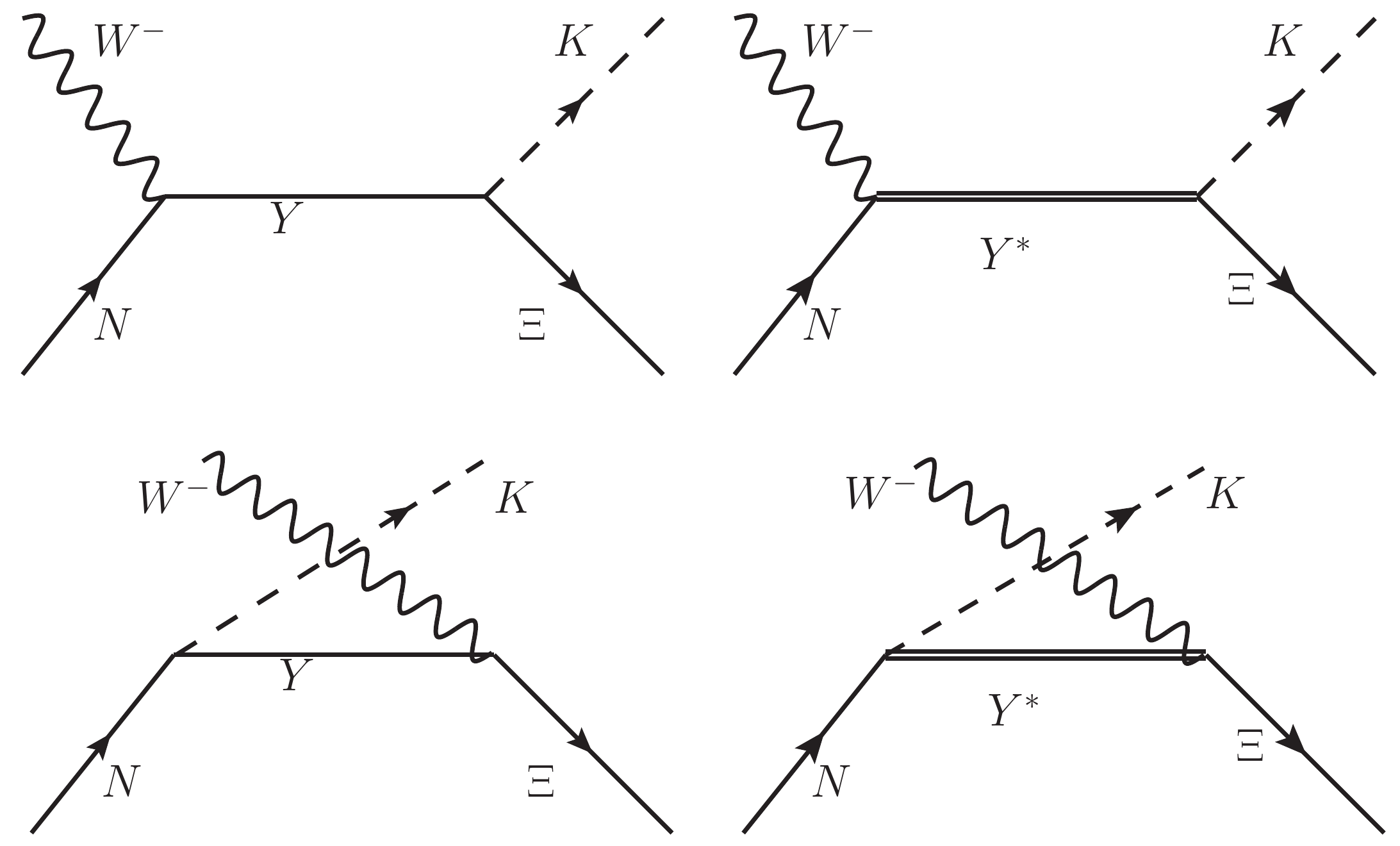}
\caption{Feynman diagrams for the 
$\Xi$ production. The intermediate 
states $Y$ are the 
($S=-1$) $\Lambda,\Sigma$ 
hyperons, and $Y^*$ is 
$\Sigma^\ast(1385)$ 
resonance.}
\label{fg:feynman}
\end{center}
\end{figure}

In Fig.~\ref{fg:feynman} we show 
the Feynman 
diagrams that contribute to the
hadronic current.  
The production mechanism  include non-resonant background (NRB) and resonance terms. 
On NRB sector the only possible choices are  $Y=\Lambda, \Sigma$ baryons. 
In resonance sector we have considered 
in this work the lowest lying resonance with strange quantum number $S=-1$, namely the $Y^\ast =\Sigma^\ast(1385)$.
This resonance state has been also previously 
considered for $|\Delta S|=1$ antikaon
production off nucleons~\cite{Alam:2011xq}. 
In the following section, we first describe 
our model for NRB followed by the implementation 
of $\Sigma^\ast(1385)$ resonance. 

\subsection{Non-resonant background}\label{NRB-section}
The NRB terms have direct and cross diagrams and the corresponding currents are 
 \begin{align}\label{eq:born_amp}
    j^\mu_{cc} \big \arrowvert_{sY} &=   
    \frac{i A_{s} V_{us}}{f_\pi}\;
    \bar u_\Xi(p^\prime_\Xi)\,  
    \spk \gamma^5  \frac{ 
    \sP + \sq +M_Y}{(p+q)^2 - M_Y^2}
    \Gamma^\mu_{N Y}\,  
    u_N(p)  \nonumber \\          
    j^\mu_{cc} \big \arrowvert_{uY} &=  
    \frac{i A_{u} V_{us}}{f_\pi}\;
    \bar u_\Xi(p^\prime_\Xi)\, 
    \Gamma^\mu_{Y \Xi} \frac{ \sP - \spk + M_Y}{(p - 
    p_k)^2 - M_Y^2} \spk 
    \gamma^5\, u_N(p) \nonumber \\ 
\Gamma^\mu_{B_i B_j}(q) 
&= f^{B_i B_j}_1(q^2) \gamma^\mu 
+ i f^{B_i B_j}_2(q^2) 
\frac{ \sigma^{\mu \nu}}{M_{B_i} + 
        M_{B_j}} q_\nu \nonumber \\ 
    &    - g^{B_i B_j}_1(q^2) 
    \gamma^\mu  \gamma^5  - 
    g^{B_i B_j}_3(q^2) q^\mu  \gamma^5 
\end{align} 
where $f_\pi$ is the pion decay constant, and 
$V_{us}(= \sin \theta_c)$ is the
corresponding  
Cabibbo-Kobayashi-Maskawa (CKM) matrix element
for weak strangeness-changing processes. 
The weak vertex function 
$\Gamma^\mu_{B_i B_j}(q)$ denotes the 
weak transition from baryon $B_i$ to $B_j$ and
it is written in terms of transition 
vector $(f^{B_i B_j}_{1,2}(q^2))$ and 
axial-vector $(g^{B_i B_j}_{1,3}(q^2))$ 
form factors. 
Experimentally, very little information is 
available regarding these form factors
for the weak strangeness-changing 
transitions between 
states of the octet baryon. Hence, 
we rely on exact SU(3) flavor symmetry 
to relate them to the
well-known proton and neutron vector form
factors ($f^{p,n}_{1,2}(q^2)$) and to the
nucleon axial-vector one ($g_A(q^2)=
g^{n p}_1(q^2)$)
by using Cabibbo's 
theory~\cite{Cabibbo:2003cu}. 
The above prescription has also been used by 
several other authors to obtain $N-Y$ and 
$Y-Y^\prime$ transitions, see for example 
Refs.~\cite{Singh:2006xp,Alam:2014bya,
Adera:2010zz}. 
In the present work we follow the formalism 
already presented 
in Refs.~\cite{Singh:2006xp,
Alam:2014bya} and we discuss it here
in brief~\footnote{For a more complete 
description please see 
Refs.~\cite{Cabibbo:2003cu,
Singh:2006xp,Alam:2014bya}.
In particular, note that in Eq.
(\ref{eq:born_amp}) we have discarded
the scalar $f_3(q^2)$ and the ``weak
electricity" $g_2(q^2)$ form factors,
both belonging to the so-called 
``second-class currents"~\cite{Weinberg:1958ut}.
For a very recent and much more
detailed discussion
on this issue and the impact of
effects from second-class currents in
some observables calculated for
quasielastic neutrino (antineutrino) production
of nucleons and hyperons, please see 
section II of Ref.\cite{Fatima:2018tzs}.}.

\begin{table}
\begin{tabular}{|c|c|c|}\hline
Transitions &$C^{B B^\prime}_a$ & 
$C^{B B^\prime}_s$ \\ \hline
$p\rightarrow p$& 1& $\frac13$ \\
$n\rightarrow n$& 0& $-\frac23$ \\
$p\rightarrow n$& 1& 1 \\
$p\rightarrow \Lambda$&  $-\sqrt{\frac{3}{2}}$&$-\frac{1}{\sqrt{6}}$\\
$p\rightarrow \Sigma^0$& $-\frac{1}{\sqrt{2}}$& $\frac{1}{\sqrt{2}}$\\ 
$n\rightarrow \Sigma^-$& $-$1& 1 \\
$\Lambda \rightarrow \Xi^- $& $ \sqrt{\frac32} $& $\frac{-1}{\sqrt{6}}$\\ 
$\Sigma^0 \rightarrow \Xi^- $& $\frac{1}{\sqrt2}$& $\frac{1}{\sqrt2}$\\ 
$\Sigma^+ \rightarrow \Xi^0 $& 1 & 1 \\
\hline
\end{tabular}
\caption{SU(3) factors 
$C^{B B^\prime}_{a,s}$ of 
Eqs.~(\ref{ff0}) and
(\ref{ff01}).}
\label{tab:su3}
\end{table}

\subsubsection{Transition form factors for non-resonant background}
\begin{figure*} 
\includegraphics[width=0.32\textwidth,height=.34\textwidth]{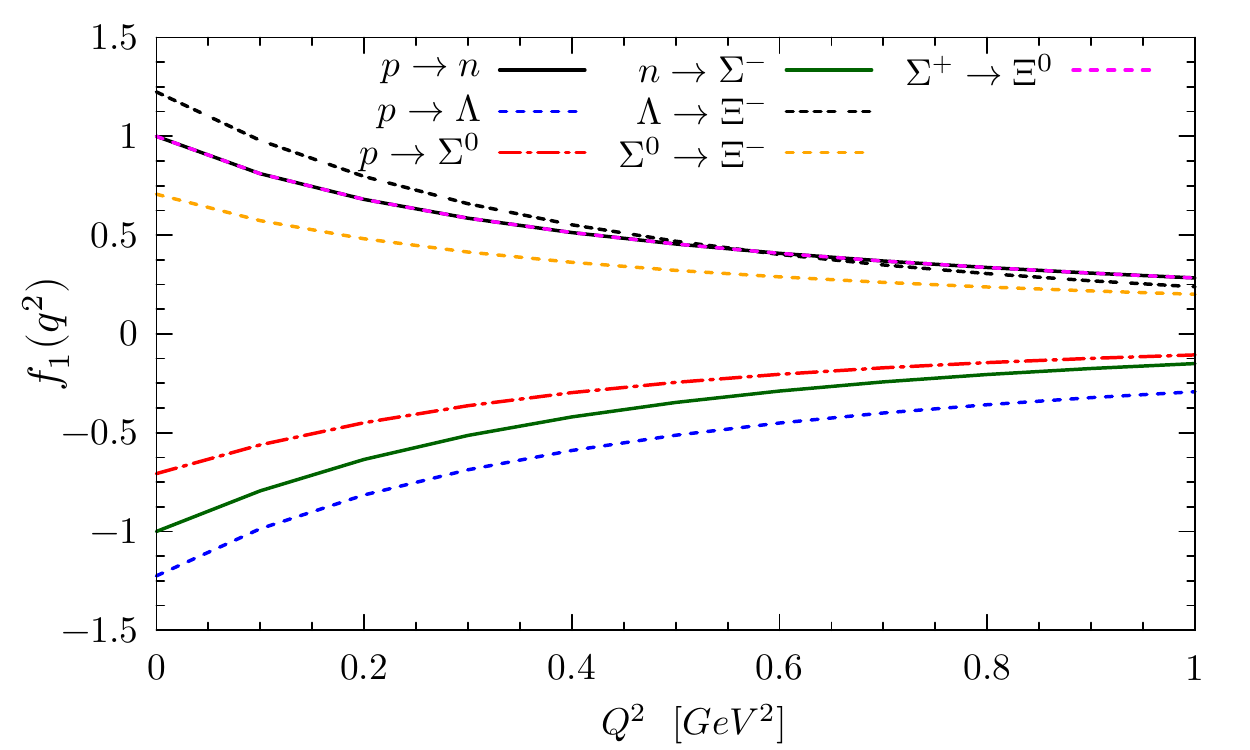}
\includegraphics[width=0.32\textwidth,height=.34\textwidth]{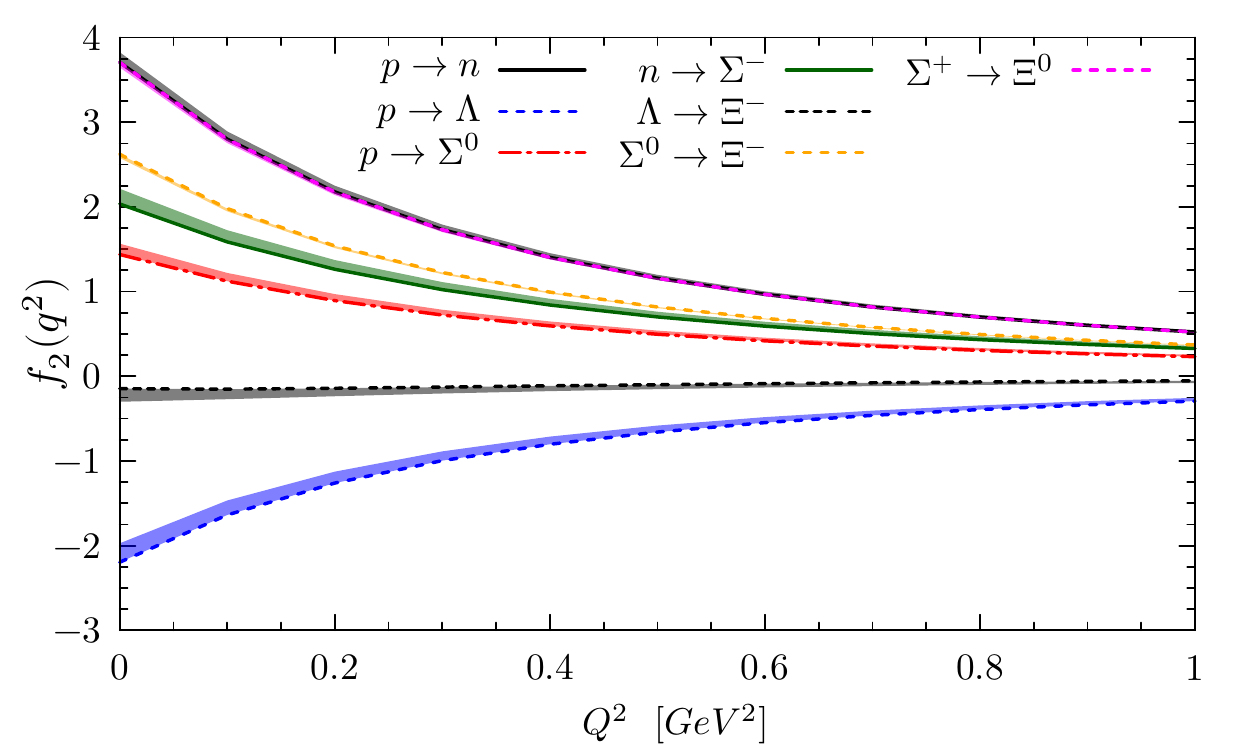}
\includegraphics[width=0.32\textwidth,height=.34\textwidth]{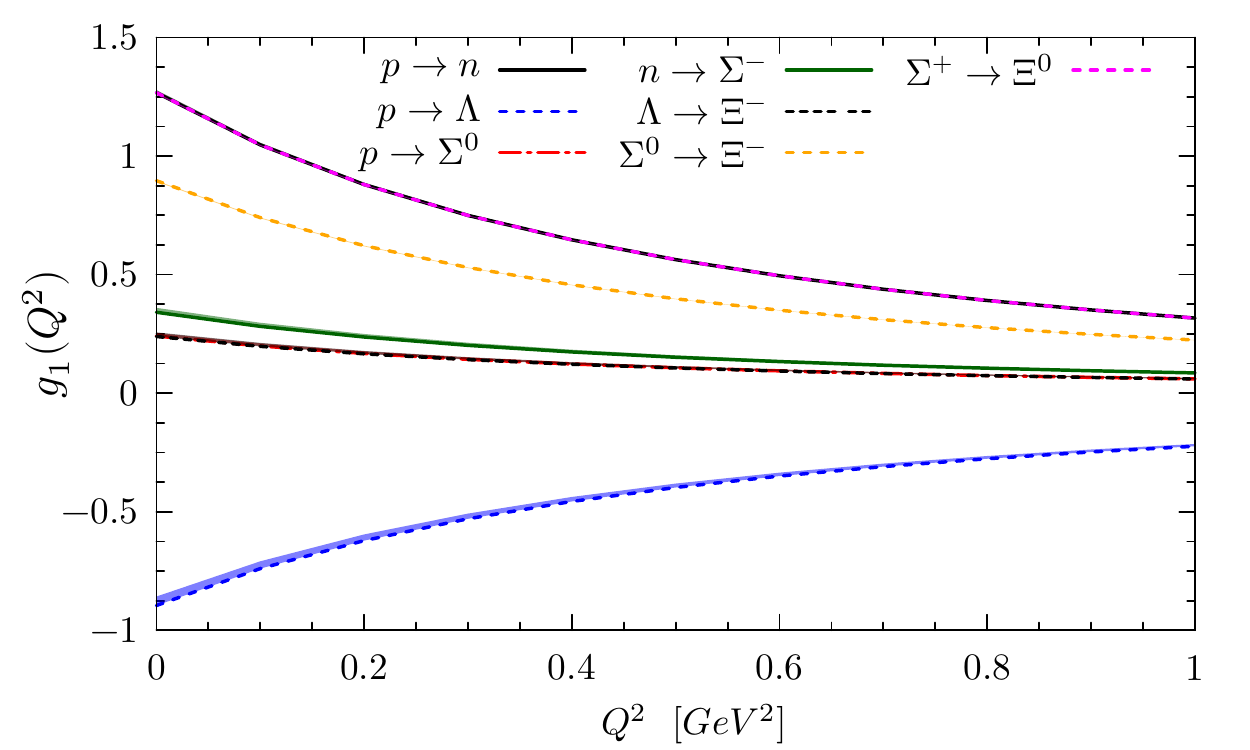}
\caption{$Q^2(= -q^2)$ dependence of
the different transition form factors. 
Shaded regions show the uncertainties 
due to SU(3) breaking.}
\label{fg:form_fac}
\end{figure*}
Assuming the octet 
representation of SU(3) in flavor 
space, one can 
relate the vector ($f_{1,2,3}(0)$) 
and axial-vector 
couplings ($g_{1,2,3}(0)$) with 
the reduced 
matrix elements corresponding to 
the antisymmetric 
$({\cal F}^{V,A})$ (also called
$f$-type) and 
symmetric $({\cal D}^{V,A})$ 
(also called
$d$-type) couplings as: 
\begin{eqnarray}
\label{ff0}
f^{B B^\prime}_i(q^2)= 
C^{B B^\prime}_a {\cal F}^{V}_i (q^2)
+ C^{B B^\prime}_s 
{\cal D}^{V}_i (q^2), \quad \text{(i=1,2)}\\
\label{ff01}
g^{B B^\prime}_i(q^2)=
C^{B B^\prime}_a {\cal F}^{A}_i (q^2)+
C^{B B^\prime}_s {\cal D}^{A}_i (q^2), 
\quad \text{(i=1,3)}
\end{eqnarray}
where $C^{B B^\prime}_{a,s}$ are related to SU(3) 
Clebsh-Gordan coefficients. 
The appearance of 
two different reduced 
transition
matrix elements 
($\mathcal{F}$
and $\mathcal{D}$) between 
octet baryon
states is because of Cabibbo 
model
\cite{Cabibbo:2003cu}, which 
assumes that the vector and 
axial-vector
currents belong to an octet 
of flavour 
currents and in the 
irreducible
representation (IR) of 
$\left\lbrace 8\right\rbrace 
\otimes
\left\lbrace 8
\right\rbrace$, the octet  
$\left\lbrace 
8\right\rbrace$ IR appears 
twice. This means
that there are two independent reduced matrix
elements when applying the Wigner-Eckart
theorem for SU(3) octet transitions through
an octet of flavor current
operators. One of these
reduced matrix elements is
related to $\mathcal{F}$ while the other one
to $\mathcal{D}$.  
 As the flavour structure for 
 both vector
(V) and axial (A) currents remains 
same, the coefficients
$C^{B B^\prime}_{a,s}$ 
given in Table \ref{tab:su3} are 
identical for both currents.
It is worth noticing that
Eqs.~(\ref{ff0}) and 
(\ref{ff01}) are generally 
expressed for the SU(3) 
couplings.  
However, it has been 
shown by previous authors~\cite{Singh:2006xp,
Alam:2014bya} that such relationships work well for finite $q^2$ too.

 The transitions $p 
 \rightarrow p$  and  
$n \rightarrow n$ can be driven
by the electromagnetic current, 
which is a linear combination of 
the hypercharge and third 
component of the isovector currents
of the octet of current operators  
in Cabibbo model 
\cite{Cabibbo:2003cu}.
Hence the corresponding
Dirac $(f_1^{p,n})$ and 
Pauli $(f_2^{p,n})$ form factors appearing
when taking matrix elements of the electromagnetic
current between proton and neutron states can also
be written as in Eq. (\ref{ff0}) with
the coefficients given in Table 
\ref{tab:su3}:

\begin{equation}\label{vem3}
\begin{aligned}
f^n_i(q^2)&=-\frac{2}{3}{\cal D}^V_i(q^2), 
\;\;\;\;\ &\text{i=1,2} \\
f^p_i(q^2)&={\cal F}^V_i(q^2)+
\frac{1}{3}{\cal D}^V_i(q^2), \;\;\;\ 
&\text{i=1,2}.
    \end{aligned}
\end{equation}

The system of 
equations given in
(\ref{vem3}) is then
inverted to 
give ${\cal D}^V_i$ and 
${\cal F}^V_i$ in terms of the well-known $f_{i}^{p,n}$ electromagnetic form factors.
In this way, all the octet transition vector form
factors $f^{BB^\prime}_i(q^2)$
can be written uniquely
in terms of the proton and neutron form factors.
The $q^2$-dependence is assumed to be driven by 
$f_{1,2}^{p,n}(q^2)$ and no additional 
$q^2$-dependence has been taken. 
We use Galster's 
parameterization~\cite{Galster:1971kv} for 
$f_{1,2}^{p,n} (q^2)$.
The variation of the different 
transition form factors with $Q^2$ are shown in 
Fig.~\ref{fg:form_fac}. The shaded 
region shows the effect due to 
SU(3) symmetry breaking, see 
Appendix~\ref{app:SU3} for 
details. We follow the 
prescription of 
Ref.~\cite{Faessler:2008ix} for 
SU(3) corrections. We found that 
among all transition form factors  
$p \to \Lambda$ transition suffers 
maximum deviation followed by $n 
\to \Sigma^-$ for both $f_2$ and 
$g_1$ form factor while $f_1$ form 
factor (for all transitions) is 
protected by Ademollo-Gatto 
theorem~\cite{Ademollo:1964sr}.  

\begin{table*}[htb]
\begin{center}
\caption{Constant factors 
($A_s$, $A_u$) in 
Eqs.~(\ref{eq:born_amp}) and 
(\ref{eq:res_amp}).}
\renewcommand{\arraystretch}{1.5}
\begin{tabular}{| l |c | c |c |  c |c |  c |c |   }\hline\hline
Process &     
\multicolumn{3}{|c|}{Direct 
term $(A_s)$ } &    
\multicolumn{3}{|c|}{Cross 
term $(A_u)$}  \\ 
        &  $Y=\Lambda$ & $Y=\Sigma$ & $Y=\Sigma^\ast$ & $Y=\Lambda$ & $Y=\Sigma$ & $Y=\Sigma^\ast$ \\ \hline
$\bar \nu_l + p \rightarrow l^+ + K^+ + \Xi^-$ &  $ - \frac{D - 3F}{2 \sqrt3 }$ &  $ \frac{D +F}{2}$ & $\frac{1}{\sqrt{6}}$
&$- \frac{D + 3F}{2 \sqrt3 }$ &   $ \frac{D-F}{2}$ & $\frac{1}{\sqrt{6}}$ \\        
$\bar \nu_l + p \rightarrow l^+ + K^0 + \Xi^0$ &  $ - \frac{D - 3F}{2 \sqrt3 }$ & $-$ $\frac{D+F}{2}$ & $-\frac{1}{\sqrt6}$
& 0 & $\frac{D-F}{\sqrt2}$ & 
$\sqrt{\frac{2}{3}}$ \\
$\bar \nu_l + n \rightarrow l^+ + K^0 + \Xi^-$ &  
0 & $  \frac{D+F}{\sqrt2}$ & $\sqrt{\frac{2}{3}}$
& $ - \frac{D+3F}{2 \sqrt3 }$ & $ - \frac{D-F}{2}$ &  $-\frac{1}{\sqrt{6}}$  \\\hline 
\end{tabular}
\label{tb:coupling}
\end{center}
\end{table*}
On the other hand, in the axial 
sector, while 
${\cal F}^{A}(0)$ and ${\cal 
D}^{A}(0)$
still form reduced 
matrix elements for antisymmetric and symmetric 
axial couplings with $C^{BB^\prime}_{s,a}$ being 
the same coefficients of Table \ref{tab:su3},  
there is no competing either electromagnetic
or weak channel which could be used to fix them
separately at finite $q^2$, 
as it was the case for their
vector counterparts in Eqs. 
(\ref{vem3}).
However, at $q^2=0$ some information is available 
from semileptonic hyperon decays 
~\cite{Cabibbo:2003cu}.
Therefore, the extraction of the axial
couplings ${\cal F}^{A}(0)$ 
and ${\cal D}^{A}(0)$ has been 
performed assuming SU(3) symmetry.
For the present work we use the numerical 
values ${\cal F}^{A}(0)=0.463$ 
and ${\cal D}^{A}(0)=0.804$ 
\cite{Cabibbo:2003cu}.

For the $q^2$-dependence of the octet
transition axial form factors, we 
first express the nucleon axial 
coupling corresponding to 
transition $p \to n$  
in terms of ${\cal F}^{A}(0)$ 
and ${\cal D}^{A}(0)$ 
and relate all of them in terms of  
${\cal F}^{A}(q^2)$ and 
${\cal D}^{A}(q^2)$:
\begin{align}
    g^{np}_1(q^2)&= 
    \frac{g_A}{(1 - q^2/M_A^2)^2} \nonumber \\ 
     &= \frac{{\cal F}^{A}(0)}{(1 - q^2/M_A^2)^2} 
     + \frac{{\cal D}^{A}(0)}{(1 - q^2/M_A^2)^2} 
     \nonumber \\ 
     &= {\cal F}^{A}(q^2) + {\cal D}^{A}(q^2).
\end{align}
In the last step 
we exploit the linear dependence of 
$g_A$ on ${\cal F}^{A}(0)$ and ${\cal D}^{A}(0)$ 
to write the
explicit dependence on $q^2$. Therefore, 
${\cal F}^{A}(q^2)$ and ${\cal D}^{A}(q^2)$ also 
have the dipole structure (this assumption
was also considered in Ref. \cite{Singh:2006xp}). 
Using Eq.~(\ref{ff01}), we obtain
the other transition 
form factors in terms of ${\cal F}^{A}(q^2)$ and 
${\cal D}^{A}(q^2)$.  The dipole parameter $M_A$ 
can then be identified as
the nucleon axial dipole mass and for 
present work we take $M_A=1.0$ GeV. 

Finally, the 
couplings $A_s$ and $A_u$ 
in Eqs.~(\ref{eq:born_amp}) are obtained from the 
SU(3) rotations at strong vertices of the
diagrams given in Fig. \ref{fg:feynman}
and are given in 
Table~\ref{tb:coupling}. It is also worth 
noticing for the reader
that we have used pseudo-vector strong
couplings at the $BB^\prime K$ vertices,
as it is obvious from the $\spk\gamma^5$
structures appearing in Eqs. 
(\ref{eq:born_amp}). We have also checked these
couplings with the expansion of Chiral 
Lagrangians at lowest order used in
Refs. \cite{RafiAlam:2010kf,
Alam:2011xq,Alam:2012ry} and found them to be
consistent.

\subsection{Resonant mechanism}
\label{Resonant-section}

The $K \Xi$ production channel may get 
contribution from the resonant mechanism as well. 
However, in absence of experimental data their 
couplings are not known. To overcome this 
difficulty we rely again on SU(3) symmetry. 
Among all the decuplet members 
only $\Sigma^\ast(1385)$ has 
the right
quantum numbers to  
decay strongly into 
$K \Xi$ channel. 
In this  section we present the model for
resonant mechanism, which is essentially
the same as in Ref.~\cite{Alam:2011xq}.

We start by writing the general expression for 
the current corresponding to  $\Sigma^\ast(1385)$ 
resonance, 
\begin{align}\label{eq:res_amp} 
    j^\mu_{cc} \big \arrowvert_{s\Sigma^*} &= i A^{\Sigma^*}_s \frac{\cal C}{ f_\pi }  \; V_{us} \;  p_k^\lambda  \;
\bar u_\Xi(p^\prime_\Xi) 
G_{\lambda \rho}  \; ( \Gamma_V^{\rho \mu}  + 
\Gamma_A^{\rho \mu} )  \; u_N(p) \nonumber \\
    j^\mu_{cc} \big \arrowvert_{u\Sigma^*} &= i A^{\Sigma^*}_u \frac{\cal C}{ f_\pi }  \; V_{us} \;  p_k^\lambda  \;
\bar u_\Xi(p^\prime_\Xi)  \; ( \Tilde 
\Gamma_V^{\mu\rho}  + \Tilde \Gamma_A^{\mu\rho} ) 
\;  G_{\rho \lambda } u_N(p) \nonumber \\
\Tilde \Gamma^{\mu \nu }_{i} & \equiv 
\Tilde \Gamma^{\mu \nu }_{i}(p^\prime_\Xi,q) 
= \gamma^0 
\left[  \Gamma^{\nu \mu }_{i}(p^\prime_\Xi,-q)  
\right]^\dagger \gamma^0, \quad i=V,A. 
\end{align}  
The parameter ${\cal  C}$ is the 
decuplet-baryon-meson strong
coupling constant appearing below in
Eq. (\ref{eq:dec_lag}).
It is a free parameter 
that can be fitted to
reproduce the $\Delta(1232)$ decay width. 
Following  Ref.~\cite{Alam:2011xq}, its numerical 
value has been taken 
as $\mathcal{C}\sim 1$\footnote{If 
it were fitted to reproduce the
$\Sigma^\ast(1385)$ partial widths, its value
would be $\mathcal{C}\sim 0.81-0.86$, thus
reflecting the amount of experimental
SU(3) breaking in nature of about 
$15-20\%$.}.

In Eqs. (\ref{eq:res_amp}),
$G^{\mu\nu}(P)$ is the Rarita-Schwinger 
spin-$3/2$ particle propagator given by: 
\begin{equation}\label{eq:su3_proj_op}
G^{\mu\nu}(P)= \frac{P^{\mu\nu}_{RS}(P)}{P^2-M_{\Sigma^*}^2+ i M_{\Sigma^*} \Gamma_{\Sigma^*}},
\qquad 
\end{equation}
where $P$ is the momentum carried by the 
resonance: $P=p+q$ for direct 
terms ($s-$channel), while 
$P=p- p_k = p^\prime_\Xi -q$ 
for cross terms ($u-$channel). 
The  operator $P^{\mu \nu}_{RS}$  may be 
identified as the 
Rarita-Schwinger~\cite{Rarita:1941mf} 
projection operator,  
\begin{widetext}
 \begin{equation}\label{eq:rarita_prop}
P^{\mu\nu}_{RS}(P)= \sum_{spins} \psi^{\mu} \bar \psi^{\nu} =- (\sPr + M_{\Sigma^*}) \left [ g^{\mu\nu}-
  \frac13 \gamma^\mu\gamma^\nu-\frac23\frac{P^\mu
  P^\nu}{M_{\Sigma^*}^2}+ \frac13\frac{P^\mu
  \gamma^\nu-P^\nu \gamma^\mu }{M_{\Sigma^*}}\right].
\end{equation}

Finally, the weak vector and axial
transition operators $\Gamma^{\alpha\mu}_V (p,q)$ 
and $\Gamma^{\alpha\mu}_A (p,q)$ in
Eqs. (\ref{eq:res_amp}) are given by
\cite{Hernandez:2007qq}
\begin{eqnarray}\label{eq:del_ffs}
\Gamma^{\alpha\mu}_V (p,q) &=&
\left[ \frac{C_3^V}{M}\left(g^{\alpha\mu} \sq -
q^\alpha\gamma^\mu\right) + \frac{C_4^V}{M^2} \left(g^{\alpha\mu}
q\cdot P- q^\alpha P^\mu\right)
+ \frac{C_5^V}{M^2} \left(g^{\alpha\mu}
q\cdot p- q^\alpha p^\mu\right) + C_6^V g^{\mu\alpha}
\right]\gamma_5 \nonumber \\
\Gamma^{\alpha\mu}_A (p,q) &=& \left[ \frac{C_3^A}{M}\left( g^{\alpha\mu} \sq -
q^\alpha\gamma^\mu \right) + \frac{C^A_4}{M^2} \left( g^{\alpha\mu}
q\cdot P- q^\alpha P^\mu \right)
+ C_5^A g^{\alpha\mu} + \frac{C_6^A}{M^2} q^\mu q^\alpha
\right],
 \end{eqnarray}
\end{widetext}
where $C_{i}^{V,A}$ $(i=3-6)$ are 
$q^2$-dependent form factors.
Their expressions are taken
directly from Ref. 
\cite{Hernandez:2007qq}
for the $n \rightarrow \Delta^{+}$ 
weak transition
with only one exception: we relate 
$C^{A}_6$ with $C^{A}_5$ by using 
partial conservation of the axial 
current (PCAC) 
assuming the coupling of the 
$W^{-}$ boson
to the $\Sigma^\ast$ 
resonance through a kaon pole in
Fig. \ref{fg:feynman},
\begin{equation}\label{PCAC_CA6}
    C^{A}_6(q^2)=C^{A}_5(q^2)
    \frac{M^2}{m^2_K-q^2}.
\end{equation}

In principle, 
our knowledge of the weak
$N\rightarrow\Sigma^\ast$
and $\Sigma^\ast \rightarrow \Xi$
transition form factors is almost none, as
these transitions cannot be driven directly by 
electromagnetic probes, 
a research field where the vast 
majority of associated production
studies \cite{David:1995pi,Janssen:2001wk,
Lee:1999kd,JuliaDiaz:2006is,Mart:2006dk,
Han:1999ck,Shyam:2009za} 
has been carried out.
However, we know that the
$\Sigma^\ast(1385)$  and 
$\Delta(1232)$ are members of same decuplet and 
we can exploit it to relate the
$\Sigma^\ast(1385)$ transition 
form factors with other relatively known ones as
those of the
$N\rightarrow \Delta(1232)$ transition 
by using again SU(3) symmetry 
arguments. 

We begin writing a Lagrangian 
describing the interaction between decuplet and 
octet baryons with meson octet: 
\begin{equation}
{\cal L}_{dec} = {\cal C} \left( \epsilon^{abc} 
\bar T^\mu_{ade} u_{\mu,b}^d B_c^e + 
\, h.c. \right),
 \label{eq:dec_lag}
\end{equation}
where $T^\mu$ is the SU(3) representation of the
decuplet fields 
and $a-e$ are the flavor indices\footnote{
For the explicit flavour realization of the
physical states of the decuplet 
fields in the Lagrangian of
Eq. (\ref{eq:dec_lag}), see for instance the
footnote 1 on Ref. \cite{Alam:2011xq}.}. 
The interaction of baryon octet 
$(B)$, decuplet and 
pseudo-scalar meson octet $(\phi(x))$ 
with external 
weak/electromagnetic currents is achieved by 
coupling left $(l_\mu)$ and right-handed 
$(r_\mu)$ external currents in the so-called
vielbein $u_\mu$:
\begin{align}
u_\mu &= i\left[u^\dagger(\partial_\mu-i r_\mu)u-u(\partial_\mu-i
l_\mu)u^\dagger\right] ,\nonumber   \\
u^2(x) &= U(x) =\exp\left(i\frac{\phi(x)}{f_\pi}\right) \,. 
\end{align}
For further details, the reader is either
referred to the review \cite{Scherer:2002tk} 
or to Refs.~\cite{RafiAlam:2010kf,Alam:2011xq},
where the same formalism has been  applied. 

In the $W N \to \Delta$ transitions 
it is well-known 
that, out of the $8$ form factors appearing in 
Eq.~(\ref{eq:del_ffs}), the most dominant 
contribution comes from $C_5^A$ (see for example 
Refs.~\cite{Lalakulich:2005cs,
Hernandez:2007qq,Hernandez:2010bx,
Alam:2015gaa}). The Lagrangian of 
Eq.~(\ref{eq:dec_lag}) only 
gives information about 
$C_5^A(0)$ for all the octet-decuplet baryons'
transitions. Thus, 
using SU(3) symmetry one can easily
relate the weak $(C_5^{A}(0))^{B\rightarrow D}$
transition couplings between baryon octet and 
baryon decuplet with only one of them taken as
reference. 
In our case we chose   
$(C_5^{A}(0))^{n\rightarrow 
\Delta^{+}}$
as a reference which has 
been extensively studied in 
past, see for example 
Ref. \cite{Hernandez:2007qq}.
However, given
the form of Eqs. (\ref{eq:res_amp}) and
(\ref{eq:del_ffs}), the SU(3) factors relating
the different weak $B\rightarrow D$ transition
vertices (given as $\Gamma^{\rho\mu}_i$
in Eqs. (\ref{eq:res_amp})) are totally hidden
in our numbers for $A^{\Sigma^\ast}_{s,u}$ of Table \ref{tb:coupling} for the different
reactions. Of course, all the other form factors
of Eq. (\ref{eq:del_ffs})
besides $C_5^{A}$ are assumed to rotate equally
under SU(3) transformations.
The same procedure and 
assumptions were made previously in 
Ref.~\cite{Alam:2011xq} for the 
$N\rightarrow \Sigma^\ast$ weak transition. 

Finally, the energy-dependent 
width ($\Gamma_{\Sigma^*}$) 
appearing in Eq.~(\ref{eq:su3_proj_op}) may be 
written as:
\begin{align}
    \Gamma_{\Sigma^*} = \Gamma_{N\bar{K}} + 
    \Gamma_{\Lambda \pi} + 
    \Gamma_{\Sigma \pi} + \Gamma_{\Xi K}
\end{align}
 where $\Gamma_{B \phi}$ is the partial decay 
 width for a decuplet $(D)$ member
 to meson $(\phi)$ and 
 baryon octet $(B)$, calculated from the decay 
 amplitude derived from the Lagrangian 
 (\ref{eq:dec_lag}), 
 \begin{eqnarray}
 \Gamma_{D \rightarrow B \phi}=\frac{C_Y}{192\pi}\left(\frac{\cal C}{f_\pi}\right)^2
\frac{(W+M_B)^2-m^2_\phi}{W^5} \nonumber \\
\lambda^{3/2}(W^2,M_B^2,m^2_\phi) \,
\Theta(W-M_B-m_\phi),
\end{eqnarray}
where $\lambda(x,y,z)=(x-y-z)^2- 4yz$ is the  
K\"{a}ll\'{e}n lambda function, $\Theta$ is the 
step function, $M_B$ and $m_\phi$
are the final baryon and meson masses, 
respectively, and $W$ is the invariant
mass carried by the resonance in the propagator
given in Eq.(\ref{eq:su3_proj_op}), i.e, 
$W^2=P^2$. Finally,
the factor $C_Y$ is 1 for 
$\Lambda\pi$ and $\frac23$ for $N\bar{K}$, 
$\Sigma\,\pi$ and $\Xi\, K$ partial
decay widths, respectively.

\section{Results}\label{sec:results}
\begin{figure}
\includegraphics[width=0.45\textwidth,height=.4\textwidth]{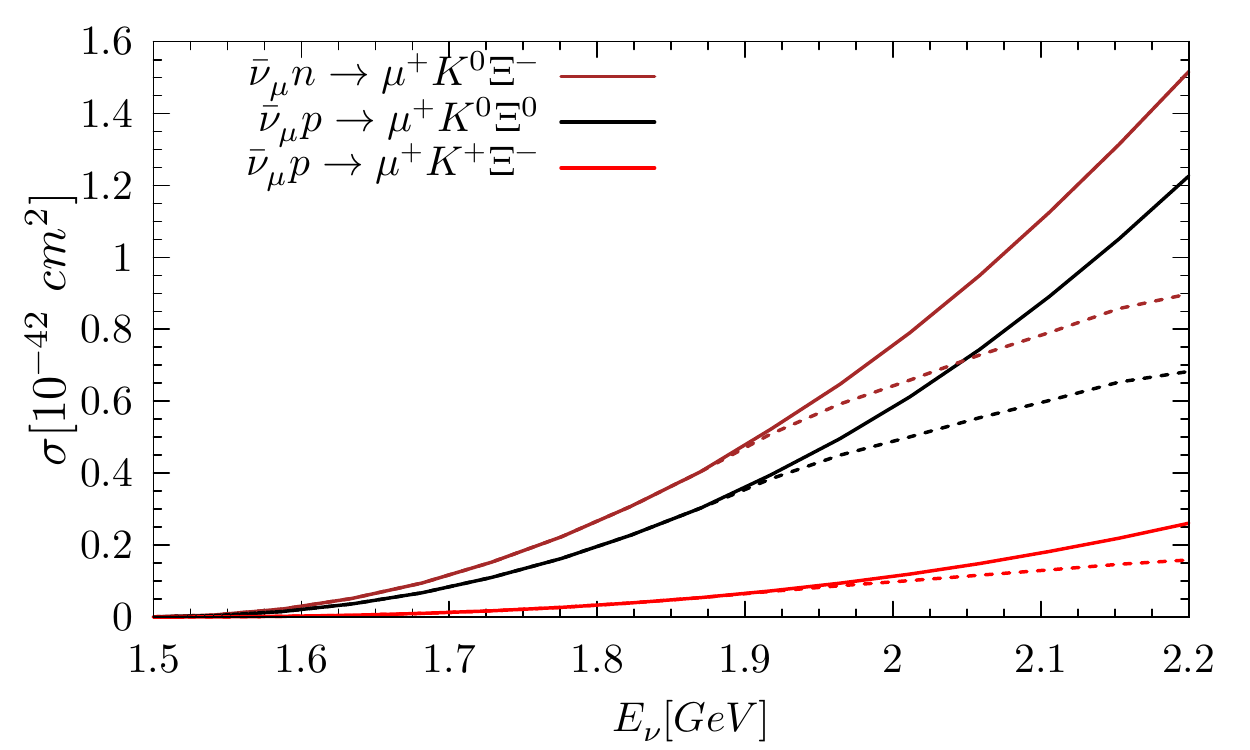}
\caption{Total cross section $\sigma$ 
vs. $E_\nu$ for the different channels of 
(\ref{eq:all_weak_ch}). 
Dashed lines show the results with 
$W_{cut}=2.0$ GeV for each process 
(same color).}
\label{fg:xsec_all}
\end{figure}
\begin{figure*}
\begin{center}
\includegraphics[width=0.32\textwidth,height=.34 \textwidth]{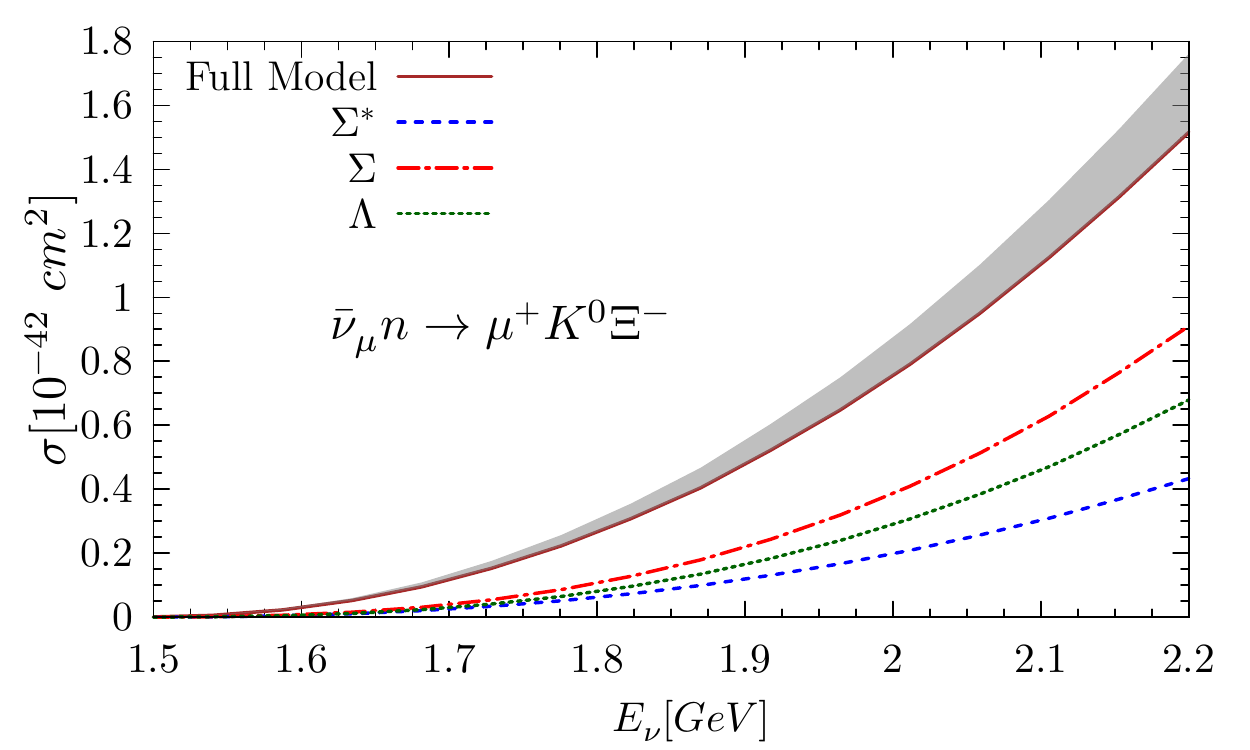} 
\includegraphics[width=0.32\textwidth,height=.34\textwidth]{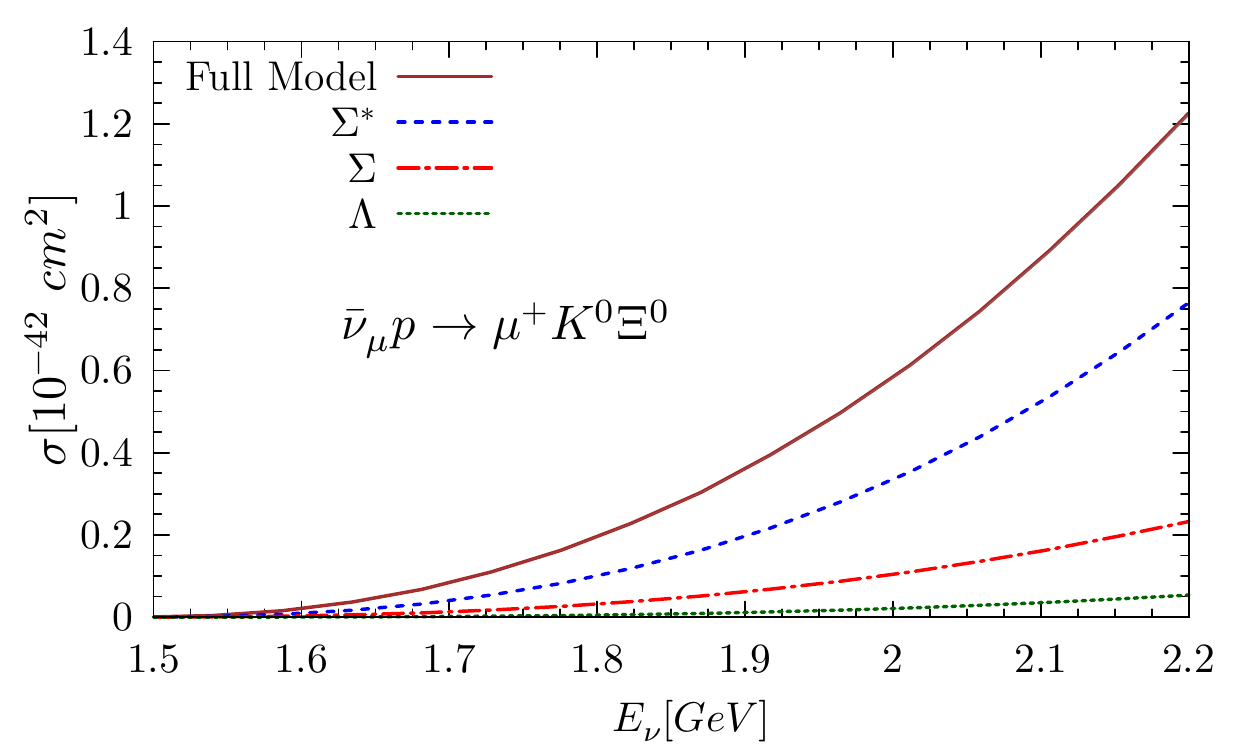}
\includegraphics[width=0.32\textwidth,height=.34\textwidth]{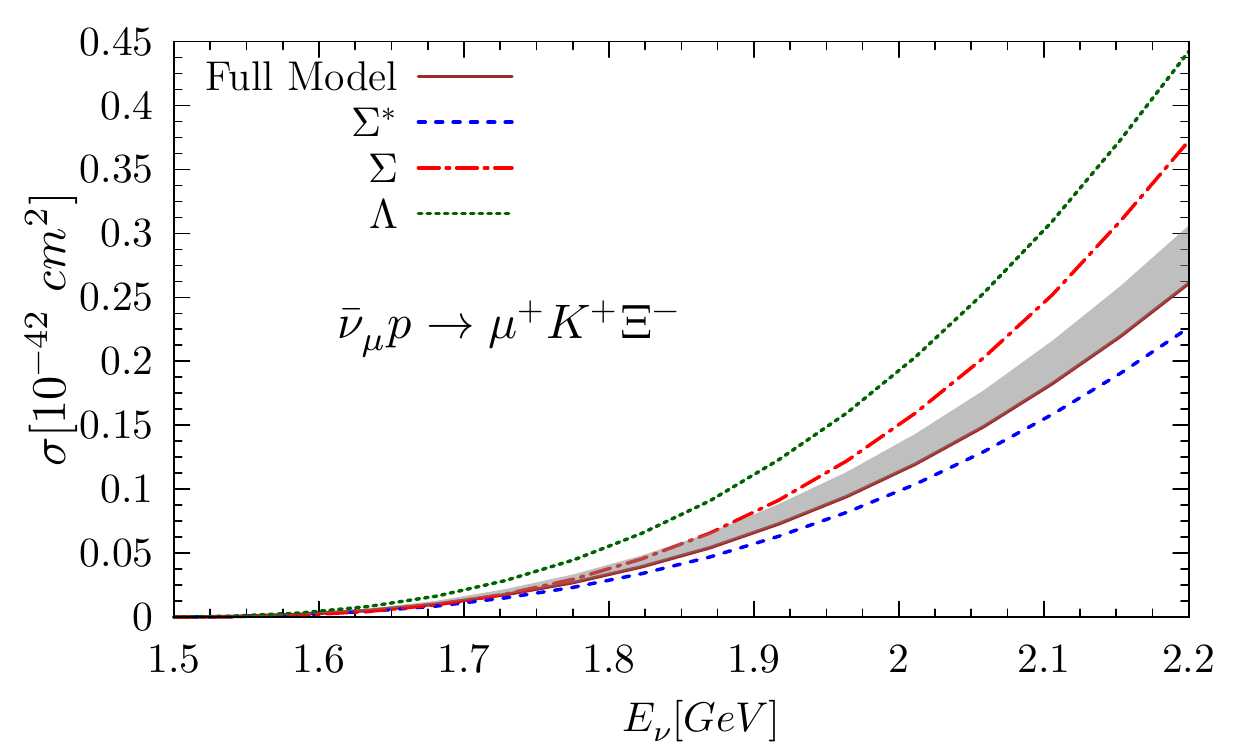}
\caption{Total cross section $\sigma$ 
vs. $E_\nu$ for the different channels of 
(\ref{eq:all_weak_ch}).
Shaded regions show 
effects due to SU(3) breaking.  }
\label{fg:all_chn}
\end{center}
\end{figure*}
The total cross sections 
corresponding to the channels given 
in (\ref{eq:all_weak_ch}) are obtained after 
integrating over $W$ and $Q^2$ in 
Eq.~(\ref{eq:diff_3b_lab_phsp}).  We present the 
results for muon anti-neutrino induced total 
cross section in Fig.~\ref{fg:xsec_all}. The
full model results are shown by solid curves, 
while dashed lines 
show the results by applying a cut in the $K\Xi$ 
invariant mass of $W_{cut}=2$ GeV  
for the corresponding 
processes (identified by the same color).  
We will discuss 
the choice of  $W_{cut}$ and the effects of it 
later in this section. 

On the other hand, considering the full model, we 
find that among the three channels $ n \to K^0 
\Xi^-$ is the most dominant one followed by 
$p \to K^0 \Xi^0$ and $p \to K^+ \Xi^-$. One 
should note that if we compare our result(s) for 
inclusive kaon production with $\Delta S=0$ 
mechanisms we find that the cross section for 
$K^0$ and $K^+$  are about 3 and 
6 percent of the 
corresponding $\Delta S=0$ processes, 
respectively. This is consonance
with the Cabibbo suppression for $|\Delta S|=1$
processes with respect to their $\Delta S=0$ 
counterparts.
For the $\Delta S=0$ kaon 
production we took the model discussed in 
Ref.~\cite{Alam:2014izc}. 
In their model,  they use only non-resonant terms
which come from lowest order expansion of chiral Lagrangian.
 However, in absence of resonant 
 mechanisms, the model 
is not reliable for high invariant mass and hence the 
comparison may be 
treated as guesstimate.

In Fig.~\ref{fg:all_chn} we present the 
contribution due to different intermediate state 
on the total cross section\footnote{For 
each term we include direct and cross diagrams. 
For example, $\Lambda$ intermediate state 
corresponds to the coherent 
sum of direct and cross 
diagrams, i.e, $j^\mu|_{s \Lambda} + j^\mu|_{u 
\Lambda}$}. Solid lines show again
the results of the full model. 
One can notice that in $\bar 
\nu_\mu n \to \mu^+ K^0 \Xi^-$ and 
$\bar \nu_\mu p \to \mu^+ K^0 \Xi^0$ the 
contributing diagrams add up, while in $\bar 
\nu_\mu p \to \mu^+ K^+ \Xi^-$ they seem to 
have strong cancellations due 
to the interferences, 
what results in a smaller cross 
section than the $\Lambda$ and $\Sigma$ term
alone. 

In fact, all three channels show different 
behavior regarding their contributing diagrams. 
Starting from $\bar \nu_\mu n \to \mu^+ K^0 
\Xi^-$, the most dominating contribution 
comes from $\Sigma$ term 
followed by $\Lambda$ and 
$\Sigma^\ast$ resonance.  The smaller 
contribution due to $\Lambda$ term can be 
understood because of absence of $s-$channel 
diagram and the larger coupling ($A_s=
\frac{D+F}{\sqrt{2}}\simeq0.896$)
of the $s-$channel $\Sigma$ current
with respect to the coupling of the $\Lambda$
term ($A_u=-\frac{D+3F}{2\sqrt{3}}\simeq
-0.63$). 
On the other hand, in this channel 
$\Sigma^\ast$ resonance has the lowest 
contribution with respect to the others. 

The other channel producing $K^0$ in the
final state is $\bar \nu_\mu p 
\to \mu^+ K^0 \Xi^0$. Unlike 
the $K^0 \Xi^-$ final state, $\Sigma^\ast$
resonance turns 
out to be the most dominating one 
followed by $\Sigma$ 
and $\Lambda$ terms. The dominance of
$\Sigma^\ast$ comes mainly through its 
$u-$channel contribution, which moreover has
a larger coupling ($\sqrt{\frac23}$ vs
$-\frac{1}{\sqrt{6}}$) than in the 
$K^0 \Xi^{-}$ case.
The smallness of the
$\Lambda$ contribution can be 
very easily understood
because of the tiny value of its coupling
in the $s-$channel ($A_s=
-\frac{D-3F}{2\sqrt{3}}\simeq0.169$),
the only occurring one. It also seems from
inspection of left and middle panels of
Fig. \ref{fg:all_chn}, and from the numerical 
values of the couplings
given in Table \ref{tb:coupling} for the
$\Sigma$ currents that the interference between
$s$ and $u$-channel diagrams is important
and destructive
for this contribution when 
both couplings have more similar values.

Finally, in the $\bar \nu_\mu p 
\to \mu^+ K^{+} \Xi^{-}$ reaction, 
it seems that, due to 
destructive interference between the 
contributing diagrams, the cross section 
gets much reduced. 
However, in this channel the
highest contribution 
comes from $\Lambda$ term and the $\Sigma^\ast$ 
resonance has the 
lowest individual contribution. 

Nonetheless, it is worth mentioning that in these
Cabibbo-suppressed associated production of two
strange particles processes, the contribution of 
$\Sigma^\ast(1385)$ resonance is truly 
important (of the same order than the NRB
terms), in contrast to the case of single
antikaon production studied in Ref. 
\cite{Alam:2011xq}, where this same resonance
was found to play a minor role. 

We have also explored the effects due to SU(3) 
breaking. For this we have 
considered  the model  of 
Ref.~\cite{Faessler:2008ix}. 
SU(3) modification in the form factors and the 
parameterization are given in 
Appendix~\ref{app:SU3}. 
The corrections have been applied 
to the form factors appearing in 
Eq.~(\ref{eq:born_amp}), 
and their effect in cross section is shown by the shaded region 
in Fig.~\ref{fg:all_chn}. 
We observe that in the 
$\bar \nu_\mu p \to 
\mu^+ K^0 \Xi^0$ channel the effect of SU(3) 
breaking is negligible, while in channels $\bar 
\nu_\mu n \to \mu^+ K^0 \Xi^-$ and $\bar \nu_\mu 
p \to \mu^+ K^+ \Xi^-$ is about 15 \%. 

At the higher neutrino energies discussed here,
due to availability of phase space, 
higher ($S=-1$) 
resonances may contribute. In particular, it 
seems that the contribution
of the $\Lambda(1520)$ can be also very 
important in the light of Ref. 
\cite{Shyam:2011ys}. However, its inclusion
in the present model is beyond the scope of
this work.

To check the validity 
of our model for the 
analyzed anti-neutrino energy range of this work,
we examine the dependence on
the invariant mass $W$. 
We restrict $W$ to remain 
slightly above the $K \Xi$ 
threshold and apply a cut in the invariant mass 
of $W_{cut}=2$ GeV in order to stay as close
to threshold as possible. The reason for this
is to minimize the effects of higher lying
resonances \cite{Shyam:2011ys} 
not included in our exploratory work.
The consequences of this cut on the total cross 
section are shown in Fig.~\ref{fg:xsec_all}. 
We observe that at $E_\nu =2.2$ GeV, 
the cross section gets 
reduced by about 40 percent for $\bar \nu_\mu n 
\to \mu^+ K^0 \Xi^-$ and $\bar \nu_\mu p \to 
\mu^+ K^0 \Xi^0$, while 20 percent  for $\bar 
\nu_\mu p \to \mu^+ K^+ \Xi^-$ process. 
Other advantages of 
using the low 
invariant mass 
kinematic cut comes 
from the NRB 
terms, because these 
have been 
obtained from the 
expansion of Chiral 
Lagrangians at 
lowest order 
\cite{RafiAlam:2010kf,Alam:2011xq,Scherer:2002tk}
and this kind of expansions are really
reliable for low energy and momentum transfers.

Finally, we have also
obtained the flux averaged total cross section 
corresponding to Minerva anti-neutrino
flux ($\Phi(E_\nu)$) 
as~\cite{Minerva,Aliaga:2016oaz},
\begin{equation}\label{eq:convolution}
\langle \sigma \rangle = \frac{\int d E_\nu \; \sigma(E_\nu) \; \Phi(E_\nu) }
{\int d E_\nu \Phi(E_\nu)} \, .
\end{equation} 

The results thus obtained are presented in Table.~\ref{tab:minerva_xsec}.
\begin{table}[H]
    \centering
    \begin{tabular}{|c|c|c|}\hline 
       Process & Without Cut & $W_{cut}=2$ GeV \\
        & $10^{-41}$ cm$^2$ & 
        $10^{-41}$ cm$^2$ \\ \hline
        $\bar \nu_\mu n \to \mu^+ K^0 \Xi^-$ & 0.795 & 0.295 \\ \hline 
        $\bar \nu_\mu p \to \mu^+ K^0 \Xi^0$ & 0.853 & 0.251 \\ \hline 
        $\bar \nu_\mu p \to \mu^+ K^+ \Xi^-$ & 0.222 & 0.076 \\ \hline 
    \end{tabular}
        \caption{Average total cross 
        section convoluted with the anti-neutrino
        Minerva 
        flux~\cite{Minerva,Aliaga:2016oaz}. }
    \label{tab:minerva_xsec}
\end{table}
We find that the 
average cross section (without invariant
mass cut) for $K^0$ production is much
higher than for $K^+$, in consonance with the
larger cross sections for $K^0$ production
(see Fig. \ref{fg:xsec_all}). 
On the other hand, 
if we observe $\Xi$ hyperon, then both 
$\Xi^0$ and $\Xi^-$ have almost similar 
occurrence. One should notice that if we apply 
$W_{cut}=2$ GeV, then the average total cross 
section gets remarkably reduced. 
Further, the kinematic suppression 
seems to affect more to 
$\bar \nu_\mu p \to \mu^+ K^0 \Xi^0$ than $\bar 
\nu_\mu n \to \mu^+ K^0 \Xi^-$. 

Although the total cross sections for the 
reactions studied in this work are normally 
between one and two orders 
of magnitude below those
studied in Ref. \cite{RafiAlam:2010kf}
for single kaon production off nucleons, 
the convoluted total cross sections
(even with the invariant mass cut) with the
anti-neutrino Minerva flux (shown in
Table \ref{tab:minerva_xsec})
are of the same order of magnitude
as those shown in Table II of 
\cite{RafiAlam:2010kf} calculated 
with different muon 
neutrino fluxes corresponding to ANL 
\cite{Barish:1977qk},
MiniBooNE \cite{AguilarArevalo:2010zc} and 
T2K \cite{Abe:2012av} experiments.

The explanation for this apparent contradiction
is not difficult to understand: the cross
sections studied in Ref. \cite{RafiAlam:2010kf}
had a quite lower energy threshold
than these for $K\Xi$ production 
($0.8$ GeV vs $1.5$ GeV, respectively), and the 
fluxes used in \cite{RafiAlam:2010kf} had a 
significant contribution also at lower neutrino
energies ($E_\nu\sim0.8$ GeV) 
and shorter high energy tails, 
oppositely to the Minerva flux, which is peaked
at $E_\nu\sim 3.5$ GeV and has a much longer
tail extending towards higher energies. 
The fluxes used in Ref.
\cite{RafiAlam:2010kf}
had their significant contribution in an energy
range where the $KN$ production reactions also had a 
sizable cross section, and they
fell off faster
than the Minerva flux. Therefore, 
at the end the 
weighted cross section given by 
Eq. (\ref{eq:convolution}) gives similar results
in both cases (Table II of Ref.
\cite{RafiAlam:2010kf} and Table
\ref{tab:minerva_xsec} of this 
work). 
This is just a consequence of 
the fact that the product 
$\sigma(E_\nu)\,\Phi(E_\nu)$ is sizable in 
an energy range where the $KN$ and the $K\Xi$
cross sections reach similar values with different fluxes. Another
reason to apply the kinematic cut $W_{cut}$
here is to avoid reaching
invariant masses far larger than $2$ GeV when
flux-averaging the cross sections with
a high energy flux such as that of Minerva 
experiment. In this way we are sure that 
the effects coming from contributing
higher resonances not
included in our model are avoided as much as 
possible, and therefore we are using our model
only in the region of invariant masses where we
think it to be more reliable.

Given the numbers of the second column of
Table \ref{tab:minerva_xsec}, and the conclusions
for the similar numbers given in Ref.
\cite{RafiAlam:2010kf}, we think that
the reactions 
(\ref{eq:all_weak_ch}) could be observed
with current measurement facilities such
as Minerva experiment.

\section{Conclusions}\label{sec:conclusion}
In this work we have
obtained the cross sections for 
Cabibbo-suppressed
associated $K \Xi$ production through charge 
current interactions induced by anti-neutrinos. 
The present work represents the 
first attempt to estimate 
the cross sections for production of
$S=-2$ baryons in weak 
processes. 
These processes are not feasible 
in neutrino mode. 

The model is a natural continuation of that
used by the same authors in Refs. 
\cite{RafiAlam:2010kf,Alam:2011xq,Alam:2012ry}.
It contains Born (NRB) terms driven by the
propagation of strange
($\Lambda$,$\Sigma$) octet
baryons, and also the lowest lying strange
resonance of the decuplet, namely the
$\Sigma^\ast(1385)$, which is found to play a 
very significant role in 
these reactions, contrarily
to the case of $\bar{K}N$ production discussed
in Ref. \cite{Alam:2011xq}. 

In a recent
work \cite{Shyam:2011ys}, a 
diagrammatically-inspired model like that 
used here has shown that the inclusion of 
higher lying strange resonances like
the $\Lambda(1520)$ is very important to
describe the 
$K^{-}p \rightarrow K\Xi$ production data. Although
the inclusion of the $\Lambda(1520)$ is out of
the scope of this article, we do not renounce
including it in future refinements of the 
present exploratory work 
in the light of the findings
of Ref. \cite{Shyam:2011ys}.

The simplicity of the present 
model makes it possible to be implemented in the 
present Monte Carlo Generators like 
GENIE~\cite{Andreopoulos:2009rq,
Andreopoulos:2015wxa}. 
The cross sections computed here 
are about one order of 
magnitude less than the corresponding $\Delta 
S=0$ associated $K$ production mechanism. 
However, they are measurable 
at the current  neutrino 
facilities like Minerva and in the proposed 
mega-detectors like Dune~\cite{Acciarri:2015uup} 
and Hyper-Kamiokande~\cite{Nakamura:2003hk}.

\section*{Acknowledgement}
This research has been 
supported by MINECO (Spain) and 
the ERDF (European Commission) grants No. 
FIS2017-84038-C2-1-P,  
FIS2017-85053-C2-1-P, 
SEV-2014-0398, and by Junta de 
Andalucia (Grant No. FQM-225). 
The authors 
would like to thank Luis 
Alvarez-Ruso and Manuel Jose Vicente Vacas for 
useful discussions. 
One of the authors (MRA) 
would like to thank the 
pleasant hospitality at University of 
Granada where part of the work has been done.  

 \appendix
\section*{Appendices}
\addcontentsline{toc}{section}{Appendices}
\renewcommand{\thesubsection}{\Alph{subsection}} 
\setcounter{equation}{0}
\renewcommand{\theequation}{A\arabic{equation}}

\subsection{SU(3) Breaking}\label{app:SU3}
Following~\cite{Ademollo:1964sr}, we write the effective Lagrangian which includes additional couplings $H_i$ $(i=1 \cdots 4)$ describing SU(3) breaking as: 
\begin{align}
{\mathcal L}_1 &= 
D \, \la \bar
B \Gamma^\mu \{\epsilon_\mu, B \} \ra \, + \, F \, \la \bar
B \Gamma^\mu [\epsilon_\mu, B ] \ra \nonumber\\[1mm]
&
+ \frac{H_1}{\sqrt{3}} \, \la \bar
B \Gamma^\mu B \{\epsilon_\mu, \lambda_8 \} \ra +
\frac{H_2}{\sqrt{3}} \, \la \bar B \Gamma^\mu
\{\epsilon_\mu, \lambda_8 \} B \ra \nonumber\\[1mm]
&+ \frac{H_3}{\sqrt{3}} \, 
\la \bar B \Gamma^\mu \epsilon_\mu B \lambda_8
  - \bar B \Gamma^\mu \lambda_8 B \epsilon_\mu \ra \nonumber\\[1mm]
& + \frac{H_4}{\sqrt{3}} \,
 \biggl(  \, \la \bar B \epsilon_\mu \ra \Gamma^\mu \la B \lambda_8 \ra
+ \la \bar B \lambda_8 \ra \Gamma^\mu \la B \epsilon_\mu \ra \, \biggr) 
\end{align}
where $B$ and $\epsilon_\mu$ are the octet of 
baryon fields and the external 
current, respectively.
Angular braces, $\la \cdots \ra$, represent the 
trace of matrices in flavor space 
and $\lambda_8$ is the $8^{th}$ 
component of the set of Gell-Mann matrices.
Finally, depending upon the nature of the
current, 
$\Gamma^\mu$ can be read as $\gamma^\mu $ or 
$\gamma^\mu \gamma^5$ for vector and
axial-vector cases, respectively. 

The  couplings of baryons are then expressed in
terms of the constants $D$, $F$ and $H_i$ as
\cite{Faessler:2008ix}:

\begin{align}\label{eq:Faessler_FCC}
{\cal F}^{np} &= D + F + \frac{2}{3} (H_2 - H_3) \,, \nonumber\\[1mm]
{\cal F}^{\Lambda p} &= - \sqrt{\frac{3}{2}}
\biggl( F + \frac{D}{3} + \nonumber\\[1mm]
 &  \frac{1}{9} (H_1 - 2 H_2 - 3 H_3 - 6 H_4)
\biggr) \,, \nonumber\\[1mm]
{\cal F}^{\Sigma^- n} &= D - F - \frac{1}{3} (H_1 + H_3) \,, \nonumber\\[1mm]
{\cal F}^{\Lambda \Xi^-} &= - \sqrt{\frac{3}{2}}
\biggl( - F + \frac{D}{3} + \nonumber\\[1mm] 
 &  \frac{1}{9} (-2 H_1 + H_2 + 3 H_3 - 6 H_4)
\biggr) \,, \nonumber\\[1mm]
 {\cal F}^{\Sigma^+ \Xi^0} &= D + F - \frac{1}{3} (H_2 - H_3) \,, \nonumber\\[1mm]
 {\cal F}^{\Sigma^0 p} &= \frac{1}{\sqrt 2} {\cal F}^{\Sigma^- n}  \,,  \nonumber\\[1mm]
  {\cal F}^{\Sigma^0 \Xi^-} &= \frac{1}{\sqrt 2} {\cal F}^{\Sigma^+ \Xi^0} 
\,.
\nonumber  
\end{align}
Note that terms proportional to 
couplings $D$ and $F$ are SU(3) symmetric, while 
couplings $H_{1-4}$ account for possible 
symmetry breaking effects.  
Explicit values of coupling 
parameters $D,F$ and $H_i$ are taken from 
Ref.~\cite{Faessler:2008ix}.  
\begin{align*}
    D&= 0.7505           \qquad & F= 0.5075\\
    H_1^{g_1} &= -0.050  \qquad & H_2^{g_1} = 0.011 \\ 
    H_3^{g_1} &= -0.006  \qquad & H_4^{g_1} = 0.037  \\
    H_1^{f_2} &=  -0.246{\mathcal X}  \qquad & H_2^{f_2} = 0.096{\mathcal X} \\ 
    H_3^{f_2} &= 0.021{\mathcal X}  \qquad & H_4^{f_2} = 0.030{\mathcal X},    
\end{align*}
where ${\mathcal X}= \frac{M+M_Y}{M}$ is a constant appearing because of different normalization convention taken for $f_2$ form factor in Ref.~\cite{Faessler:2008ix} and in present case (see Eq.~(\ref{eq:born_amp})). 
Further assumptions taken are:  
\begin{enumerate}
    \item The vector coupling 
    $f_1(0)$ does not receive any 
    correction due to SU(3) breaking 
    because it is protected against these
    effects by the Ademollo-Gatto 
    theorem~\cite{Ademollo:1964sr}. 
    \item Second class currents
    ($f_3, \, g_2$), which have opposite sign 
    under $G-$parity transformation if
    compared to first class 
    currents ($f_{1,2},  \, g_{1,3}$), are 
    ignored.    We show 
    only first class currents in 
    Eq.~(\ref{eq:born_amp}). 
    
    \item For $q^2$ dependence of 
    the breaking parameters 
    ($H_i^{g_1,f_2}$), we assume 
    a dipole form with dipole 
    parameter taken as $M_A(M_V)$ 
    for axial(vector) form factors,
    respectively.  
    
    \item We assume that the PCAC and 
    Goldberger-Treiman relation is valid even 
    when $g_1$ receives SU(3) corrections. 
    Further, $g_3$ will get modified 
    from $g_1$.  In particular,
    PCAC and kaon-pole dominance implies
    the substitution $g^{B_i B_j}_3(q^2)
    \rightarrow - g^{B_i B_j}_1(q^2)
    \frac{\sq}{q^2-M^2_K}$
    in eq. (\ref{eq:born_amp}). This
    replacement has been made for all the
    calculations presented in this work.  
\end{enumerate}

Finally, care should be taken 
while using the values for 
$D$ and $F$. In our numerical calculations we 
used non-breaking couplings as $D=0.804$ and 
$F=0.463$ \cite{Cabibbo:2003cu}. However, in order to be consistent 
with Ref.~\cite{Faessler:2008ix} we use their 
parameters for SU(3) breaking
calculations.

\bibliography{biblio}
 
\end{document}